\documentclass[trackchanges,twocolumn,twocolappendix]{aastex701}
\usepackage{newtxtext,newtxmath}
\usepackage[T1]{fontenc}
\DeclareRobustCommand{\VAN}[3]{#2}
\let\VANthebibliography\thebibliography
\def\thebibliography{\DeclareRobustCommand{\VAN}[3]{##3}\VANthebibliography}
\renewcommand{\thesection}{\arabic{section}}
\usepackage{graphicx}
\usepackage{amsmath}
\usepackage{dcolumn}
\usepackage{bm}
\usepackage{booktabs}

\newcommand{\vDLeff}{\vec D_{L,\text{eff}}}
\newcommand{\DLeff}{D_{L,\text{eff}}}

\begin{document}

\newcommand{\leu}[1]{\textcolor{purple}{\textit{LEU: #1}}}
\newcommand{\edit}[1]{\textcolor{blue}{\textit{#1}}}

\title{Localization of gravitational-wave source behind galaxy clusters by gravitational-wave lensing}

\author{Brian H. Y. Cheng}
\affiliation{Department of Physics, The Chinese University of Hong Kong}
\email[show]{hycheng@link.cuhk.edu.hk}  

\author{Laura Uronen} 
\affiliation{Department of Physics, The Chinese University of Hong Kong}
\email{laura.uronen@gmail.com}

\author{Souvik Jana} 
\affiliation{Department of Physics, The Chinese University of Hong Kong}
\email{souvik.jana@cuhk.edu.hk}

\author{Samuel Lange} 
\affiliation{Department of Physics, The Chinese University of Hong Kong}
\email{samuel.c.lange@cuhk.edu.hk}

\author{Thomas E. Collett}
\affiliation{Institute of Cosmology and Gravitation, University of Portsmouth, Portsmouth PO1 3FX, UK}
\email{thomas.collett@port.ac.uk}

\author{Otto A. Hannuksela}
\affiliation{Department of Physics, The Chinese University of Hong Kong}
\email{hannuksela@phy.cuhk.edu.hk}

\begin{abstract}
\noindent
Gravitational-wave observations have opened new opportunities to study massive 
object events like merging binary black holes. 
When combined with 
electromagnetic counterparts, accurate localization of gravitational-wave 
sources could allow us to identify their host galaxies and carry out follow-up
studies, such as study the environment of merging black holes and  determine
the Hubble constant. 
One of the possible methods 
for precise localization is strong gravitational-wave lensing. 
Previous studies 
have considered strong gravitational-wave lensing with galaxy lenses. 
However, galaxy clusters have yet to be investigated due to the complexity of their mass structures. 
Here we take the list of known cluster lenses from the Hubble Frontier Fields, 
simulate lensing properties of gravitational waves from their reconstructions as they would be observed by the ground-based 
gravitational-wave detectors. When same model family is used in simulation and localization, we find that the correct lens is identified for around 90\% of the simulated events.
We also show that the source positions within 
the cluster can be pinpointed at sub-arcsecond precision  with an estimate on source redshift. When we marginalize over all possible lens models for the clusters, the lens identification and localizations remain valid as long as the correct lens model is contained in the available cluster lens models. However, most localization attempts fail at initialization if the correct cluster model is excluded.
These tests establish a proof of concept of localization under idealized conditions, while more work is needed to understand the cluster lens modeling discrepancies and reconstruction bias to validate the approach further.
\end{abstract}


\section{Introduction}
\label{sec:intro}
Observations of gravitational waves (GWs) from binary compact objects provide a
powerful tool for gravitational research. Analysis of GW signals allows us to
test gravitational theories and directly detect electromagnetically
invisible sources, such as binary black holes
(BBHs)~\citep{PhysRevD.112.084080}.

For cosmological measurements such as the Hubble
constant~\citep{AbbottR2017Agss}, GW observations can provide strong
constraints if the sources are well localized. Identifying their host galaxies
further enables multi-messenger observations of GW and electromagnetic (EM) signals from local
environments~\citep{2010PhRvD..82j2002N}. Such studies support independent distance measurements without
relying on indirect methods such as the distance
ladder~\citep{AbbottR2017Agss}, highlighting the essential role of GW
localization.

However, localizing GW sources remains a major challenge. Hundreds of
GW events have been identified by the LIGO–Virgo–KAGRA (LVK) detector networks.
The number of detected GW events is expected to rise substantially in future LVK observing runs~\citep{O3-5_prospects} and increase by orders of magnitude with planned third-generation ground-based GW observatories, such as the Einstein Telescope and Cosmic Explorer~\citep{Maggiore_2020,evans2021horizonstudycosmicexplorer}. However,
most current LVK network localizations still span tens to hundreds of square degrees, enclosing large numbers of galaxies~\citep{HannukselaOttoA2020Lmbh}, even though some well-measured events can be localized much more tightly (for
example, GW190814 was localized to $18.5\,\mathrm{deg}^{2}$ at 90\% credibility~\citep{LIGOScientific:2020zkf}). Without an EM counterpart,
associating GWs with their host galaxies is extremely difficult, particularly for
BBHs, which do not emit light. One of the few exceptions is GW170817, the
binary neutron star merger that produced a number of EM counterparts including
a gamma-ray burst, allowing better estimation of its sky
location~\citep{AbbottR2017Agss}. By contrast, GW events by BBHs still face the
challenge of poor localization.

Strong gravitational lensing of GWs may be a solution to the problem. Like EM
waves, GWs are bent and magnified when passing near massive objects such as
galaxies or galaxy clusters, producing effects such as delaying and amplifying GW signals~\citep{1998Lensing}. Strong lensing, in particular, can create multiple copies of the same signal with different arrival times and amplitudes, and potentially distorting the signals in a frequency-dependent manner~\citep{Takahashi_2003}. Previous studies have shown that if a GW is
strongly lensed, the possible sky location of the source can be greatly
reduced~\citep{Janquart_2021}, and the candidate host galaxies can be narrowed
to one or a few using the ratio between time delays and
magnifications~\citep{HannukselaOttoA2020Lmbh,Wempe2024OnThe}.

Previous GW lensing studies have focused on galaxy
lenses~\citep{HannukselaOttoA2020Lmbh, 2025RSPTA.38340152U}, while almost all
such studies exclude detailed analyses of cluster lensing. The reason is that
complex and irregular mass distributions make cluster lenses challenging to
model analytically~\citep{2017MNRAS.472.3177M, Sengul:2023olf}.

However, cluster lensing has unique advantages over galaxy lensing in both
GW-only and multi-messenger studies. Complex cluster mass structures allow
greater image multiplicity than typical smooth lens
models~\citep{Menghetti2007}. While common galaxy-lens models, which are
usually smooth mass models, produce a maximum of four images, cluster lenses
can produce more. A lensed GW signal with larger number of images has more constraints to cross-match the GW image properties against the EM lens reconstructions.
Also, the images can
impose stronger constraints on the sky locations of GW sources in parameter
estimation (PE).
The association
of lensed GW signals with cluster lenses is also easier. Galaxy clusters are much larger than galaxies and are therefore easier to observe.
Moreover, there are fewer candidate lenses in a sky patch where the lensed signals are
localized using GW data only.
The distinctive shapes of clusters can also limit the possible candidates for a detected lensing signal.
With a multi-messenger association, it may also be possible to estimate the source redshift.
This quantity is closely tied to lensed GW signals, as described in the next
section.
Therefore, cluster lenses should be included for a comprehensively study on how strong lensing affects GW source localization.

Although no lensing signals have been confirmed in current
observations~\citep{Hannuksela_2019,Abbott_2021,Janquart_2023,LIGOScientific:2025cwb},
the rate of lensing events in O5 is expected to be around 1
$\mathrm{yr}^{-1}$~\citep{WierdaA.RenskeA.C.2021BtDH,Smith_2023,2024arXiv240707526P}. Cluster lensed events may be more difficult to detect because of the larger time delay allowing higher fraction of false coincidences~\citep{Hannuksela_2026}. However, the
third-generation GW detectors are expected to detect orders of magnitude
more signals~\citep{cosmicexploreruscontribution,Maggiore_2020,evans2021horizonstudycosmicexplorer,2024arXiv240707526P} and hence more
lensed signals, making cluster lensed events detections increasingly promising in future observing runs. In this study, we ask whether the detection of multiple images of a cluster-lensed GW signal can identify the cluster lens and constrain the position and redshift of the source. This extends previous work on localization with galaxy lenses to galaxy clusters.

In this paper, we first describe the detailed method to study localization of
cluster-lensed GW signals by simulations in Sec.~\ref{sec:method}, then present
the localization results of the simulated events in Sec.~\ref{sec:result}.
Finally, we conclude in Sec.~\ref{sec:conclude} with discussions on the
results, design and potential extensions of the study.

\section{Method}
\label{sec:method}
Instead of analytical models, we adopt published numerical galaxy cluster lens reconstructions available in the
literature and build the GW localization framework on top of these models. In particular, using selected numerical lens models (Section~\ref{ssec:numerical_models}), we simulate GW events with a given position in the
source plane $(\beta_x, \beta_y)$ and a source redshift $z_s$ and obtain the lensing observables: the effective luminosity distances $\vDLeff$ and the relative time delays $\Delta \tau_k$ between image $k$ and the first image (Section~\ref{ssec:simulation}). Next, we input these observables into the localization algorithm for one of the clusters (Section \ref{ssec:evaluation}), which maximizes the likelihood\footnote{we assume the Laplace's approximation for the gravitational-wave likelihood, which is explained in Section~\ref{ssec:evaluation}} 
of producing images with the observed delays and distances.
We repeat this process for each cluster lens candidate, then determine the most probable lens using an approximation of posterior odds.

We assume throughout that the true cluster is among the candidate lenses. We use the same lens reconstruction for simulation and recovery as an idealized test of the localization framework. We also perform localization tests marginalizing over all available models of the clusters to verify the framework when there are multiple models of the same cluster with discrepancies. Finally, we test recovery with the generating reconstruction excluded for robustness tests when available models may not adequately recover the detail lensing properties of the cluster.

\subsection{Numerical Galaxy Cluster Models}
\label{ssec:numerical_models}
Current analytical models of galaxy clusters, such as Navarro–Frenk–White
profiles~\citep{Navarro_1997} and singular isothermal sphere (SIS)-like
models, can fail to capture realistic cluster mass structures due to the
complicated lensing properties in clusters, such as substructures and
asymmetries~\citep{Meneghetti_2003_lensing}. 

Therefore, we use existing numerical models for known
cluster lenses to probe our method while retaining realistic cluster lens complexity. The numerical cluster lens models used in this study are from the Hubble Space Telescope Frontier Fields (HSTFF) program~\citep{HSTFF}. We apply the numerical lens models from the Clusters As
Telescopes (CATS) team~\citep{jauzac:hal-01056116,Jauzac_2016,cats_richard_2014}, Keeton~\citep{2014MNRAS.443.3631M} and Sharon~\citep{Johnson_2014}. The idealized test adopts CATS reconstruction for each cluster, while other tests make use of all the reconstructions. Details of such tests are shown in Sections~\ref{ssec:model_marginalization} and~\ref{ssec:mismatch_design}.

\begin{figure}
\includegraphics[width=\columnwidth]{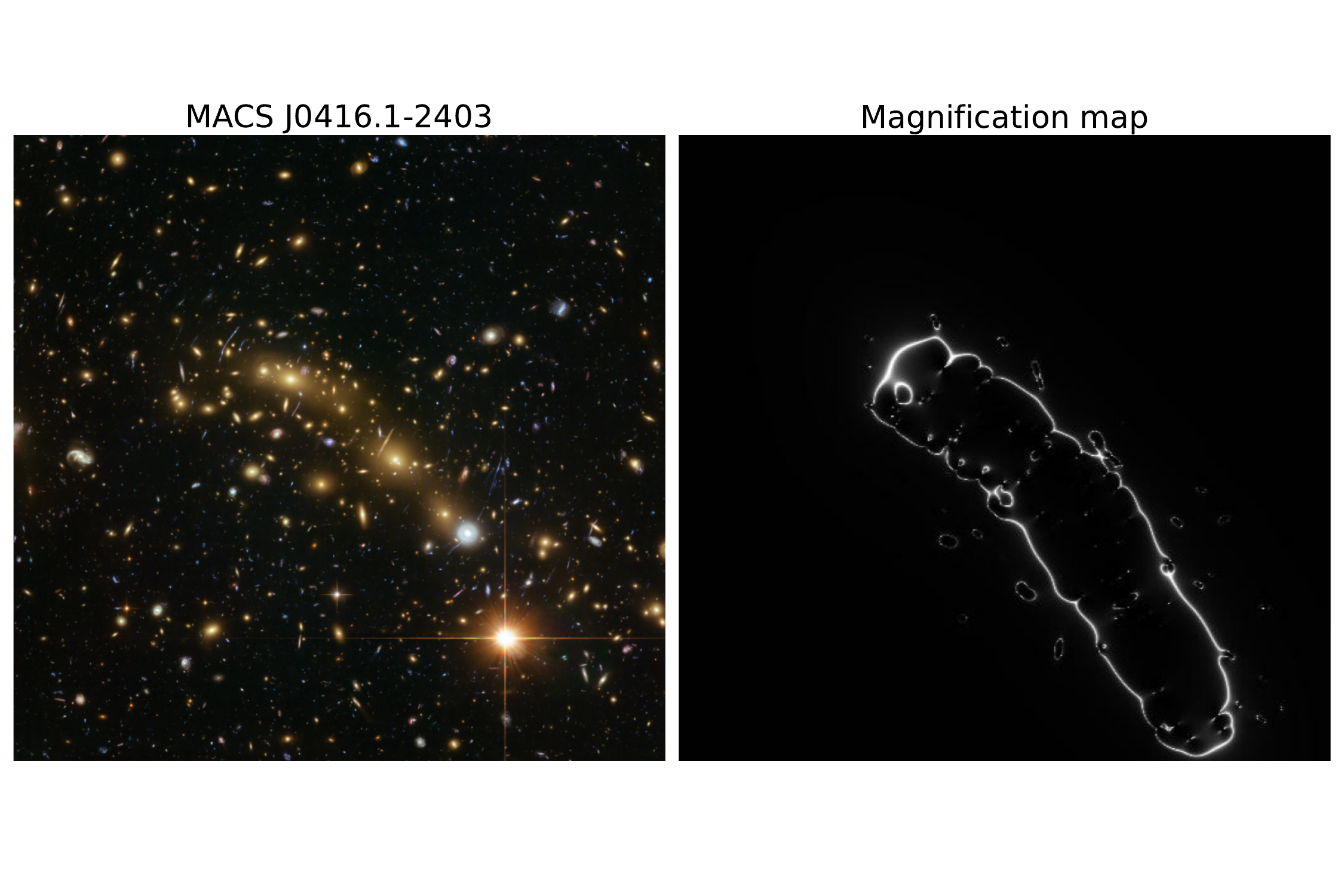}
    \caption{
	Full-depth image of the galaxy cluster MACS J0416.1-2403 (left, imaged
	with HST~\citep{HSTFF}) and its magnification map (right). The
	full-depth image shows the distribution of galaxies within the cluster
	and images produced by the cluster, and the magnification map shows the
	model caustic indicated by white curves. The irregular magnification pattern illustrates the complex mass distribution of the cluster.
	}
    \label{fig:illu}
\end{figure}

Fig.~\ref{fig:illu} shows a full-depth image of the galaxy cluster MACS
J0416.1-2403 and its numerical model of magnification map as an illustration of
the cluster structure.

The numerical lens models~\citep{Jauzac_2016,cats_richard_2014} consist of deflection-angle maps and lens-potential maps, which allow the calculation of image positions given a source position.
The details of the calculation for
model deflection angles $\vec\alpha$ and the uncertainty of the model are presented in Appendix~\ref{app:maps}.

\subsection{Simulation of Lensed GW Event Parameters}
\label{ssec:simulation}

In order to test the localization approach with expected observables of cluster-lensed signals, we simulate lensed GW events using the numerical lens models and a set of
simulated source position and redshift parameters $(\beta_x, \beta_y, z_s)$, then extract the required simulated data for injection.
GW sources are introduced as point sources in this study because the source size is negligible
compared with the lensing scale.
Given the source positions, the image positions of a lensed GW event can be obtained directly using the lens equation~\citep{Schneider1992}:
\begin{equation}
    \vec{\beta} = \vec{\theta}- \vec{\alpha}(\vec{\theta}),
\end{equation}
where $\vec{\theta}$ is the image position, $\vec{\beta}$ is the source
position and  $\vec{\alpha}(\vec{\theta})$ is the deflection angle at the image
position. Therefore, we can calculate the image positions with the source
position and deflection angle map by ray tracing.

The numerical models
include maps that contain the lens potential $\psi(\vec{\theta})$ at each
pixel.
The time delays between an unlensed GW signal and its lensed images can
then be calculated as
\begin{equation}
    \tau(\vec{\theta},\vec{\beta}) = D_{\Delta t}(\frac{1}{2}|\vec{\theta} - \vec{\beta}|^2 - \psi(\vec{\theta}))\,,
\end{equation}
where $\tau(\vec{\theta},\vec{\beta})$ is the arrival-time function in terms of Fermat potential and $D_{\Delta t}$ is given by
\begin{equation}
    D_{\Delta t}=\frac{1+z_l}{c}\frac{D_l D_s}{D_{ls}}\,,
    \label{eq3}
\end{equation}
where $D_l, D_s, D_{ls}$ are the angular diameter distance of the lens, the source, and between the lens and source respectively.

For actual observations, the relative time delays between images $\Delta
\tau_k$\footnote[1]{$\Delta \tau_k = \tau(\vec{\theta_k}|\vec{\beta}) - \min(\tau(\vec{\theta_j(\beta, z_s)}|\vec{\beta}))$, where
$\vec{\theta_j(\beta, z_s)}$ is the set of images from a GW source.} are
considered instead.
Note that $D_{s}$ scales with $z_s$, thus changing $z_s$
will significantly change the value of $\Delta \tau_k$.

In addition to the relative time delays, PE of the GW signals can also recover
the effective luminosity distance of the source. The luminosity distance $D_L$
is related to $z_s$ by:
\begin{equation}
    D_L = (1+z_s) \frac{c}{H_0} \int_0^{z_s} \frac{dz}{\sqrt{\Omega_m(1+z)^3+\Omega_\Lambda}}\,,
    \label{eq5}
\end{equation}
where we fix $H_0=70\, \mathrm{km/s/Mpc}$ in this study. For lensed events, the
observed luminosity distance of image i is equivalent to
the effective luminosity distance $\DLeff^i$ due to the signal
amplification by lensing:
\begin{equation}
    \DLeff^i = \frac{D_L}{\sqrt{|\mu_i|}}\,,
\end{equation}
where $\mu_i$ is the magnification of corresponding image. Therefore, the
observed luminosity distances can act as extra constraints for the estimation
of $z_s$ in localization, and we include these observables in the localization
algorithm.

For each source, we use the Python package $Lenstronomy$~\citep{BirrerSimon2015GLMW} to compute the relative arrival time delays and effective luminosity distances of its images. We use these values as zero-noise injection.

We use the $Lenstronomy$~\citep{BirrerSimon2015GLMW} Python package to
calculate lensed images with numerical models, source position and $z_s$.
Relative time delays and effective luminosity distances can then be calculated
by $Lenstronomy$ and used as injections for localization.

Here, only $\Delta \tau_k$ and $\vDLeff$ are simulated as
lensed GW observables.
After simulation, we treat $\Delta \tau_k$ and
$\vDLeff$ as the results of GW signal PE and then inject them into the
localization algorithm to construct a mock-data study.
The uncertainties of observables due to the lens model uncertainties are accounted in the localization.
We also assume that $z_l$ has already been measured by EM observations and is included in the numerical models.

\begin{figure}
    \includegraphics[width=\columnwidth]{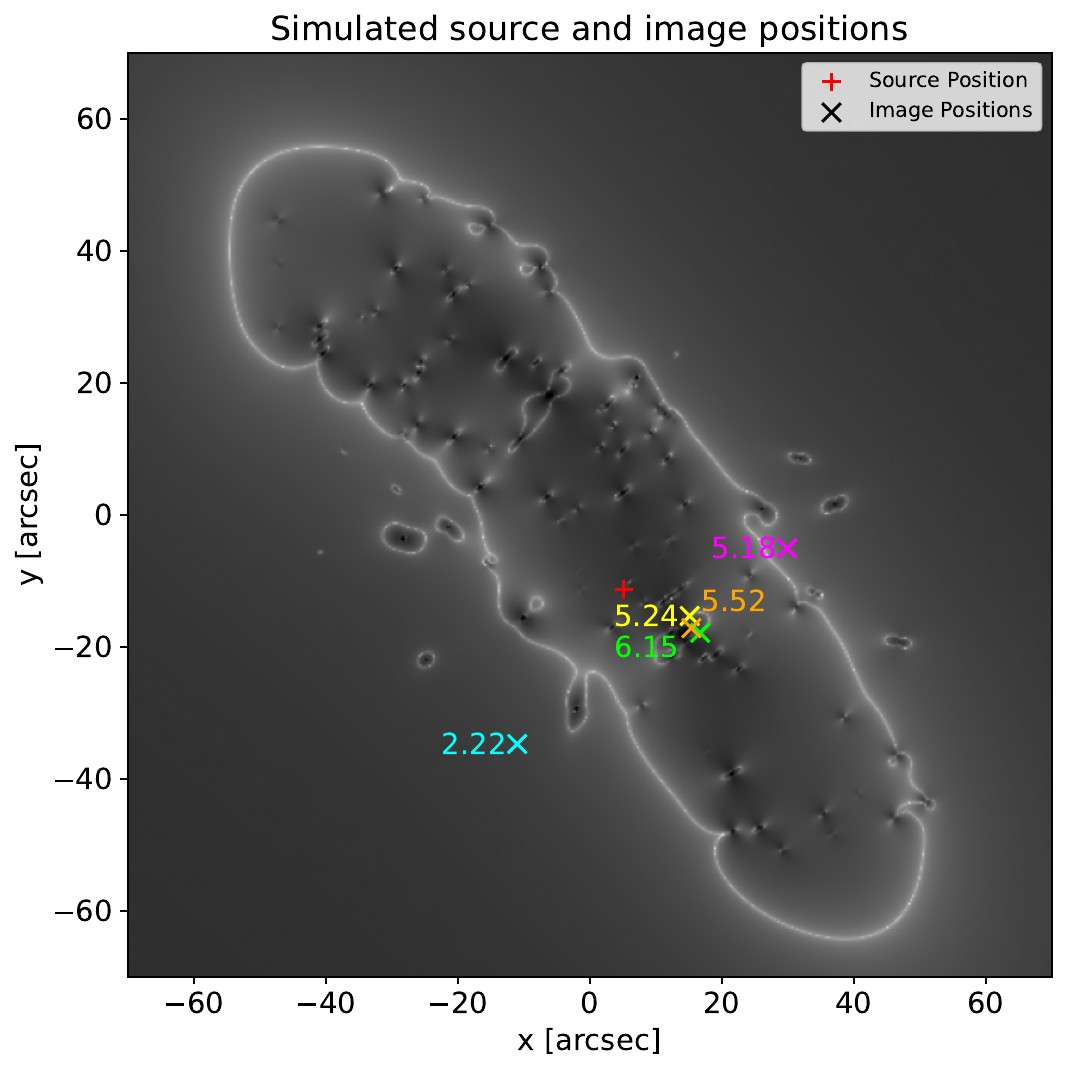}
    \caption{
	Example of a five-image lensing event lensed by the galaxy cluster MACS J0416.1-2403 with its critical curves shown. 
	The white curves represent critical curves. 
	The red plus and the crosses mark the simulated source position in source plane and its corresponding images in image plane, respectively. 
	The numbers next to the crosses indicate the absolute magnification. 
	When the source is located inside the inner caustic and near the complex mass distribution, more images can be produced beyond the typical maximum number of images of analytical galaxy lens models which is 4.
	}
    \label{fig:five_image_gw_illustration}
\end{figure}

We simulate lensed GW events from the numerical lens
models of 6 galaxy clusters and assume that the given clusters are all the
possible candidate lenses of the lensed GW events.
In the main analysis, we generate parameter sets that produce 5-image
lensed GW events (yielding $4 \,\Delta \tau_k$ and 5 $\DLeff$ from the observations).
We simulated a total of 60 5-image lensed GW events, with 10 events behind each galaxy cluster.
We consider five-image GW configurations as a high-information scenario to investigate the constraining power of cluster lensing specifically.
A large number of images is possible in cluster lensing due to the complex lensing environment of galaxy clusters.
An example of a 5-image lensed GW event is shown in Fig.~\ref{fig:five_image_gw_illustration}.

The high-multiplicity lensing configurations are not restricted to idealized mass distributions and have been observed electromagnetically.
For example, the galaxy cluster SDSS J2222+2745 produces six observed images~\cite{Dahle_2013}.
These systems demonstrate that galaxy clusters can generate image multiplicities of five or more.
Note that the detectable frequency of such systems depends on the image magnifications, source population and detector sensitivity, which are not evaluated in this work.

For the marginalized localization analysis, we select 18 of the 60 simulated five-image events. For the model-mismatch test, we simulated five 4-image events (yielding $3 \,\Delta \tau_k$ and 4 $\DLeff$ from the observations) for each model used for each of the 6 clusters. The simulation of 4-image events is to avoid larger differences between reconstructions for high-multiplicity configurations.

\subsection{Localization and Evaluation}
\label{ssec:evaluation}

We perform the localizations by sampling the source positions and redshifts under each detected cluster lens candidate so as to obtain the localization accuracy in general.
We do so using the same numerical lens model as in the simulation and assume that the CATS model is adequate to describe the lensing structure for a constrained study in the main analysis.

For a candidate cluster $i$, we denote the source position and redshift parameters by $\beta, z_s$.
The GW data $d_\text{gw}$ includes the relative arrival-time delays $\Delta\tau_k$ and the observed effective luminosity-distances of the associated images $\vDLeff$, while $d_\text{em}^i$ denotes the EM lens model of candidate $i$ and the full EM
candidate catalog is $d_\text{em}=\{d_\text{em}^1,d_\text{em}^2,\ldots,d_\text{em}^M\}$, where $M$ is the number of detected cluster candidates.
The posterior for candidate $i$
and its source parameters, assuming the GW is lensed and the lens is in the detected EM candidate set, is then
\begin{equation}
\begin{split}
&p(i,\beta,z_s|d_\text{gw},d_\text{em},H_\text{gw}^\text{L},H_\text{em}^\text{det})\\&=
\frac{
p(d_\text{gw}|\beta,z_s,d_\text{em}^i,H^i,H_\text{gw}^\text{L})
p(\beta,z_s|d_\text{em}^i,H^i,H_\text{gw}^\text{L})
\pi_i^\text{det}
}{
p(d_\text{gw}|d_\text{em},H_\text{gw}^\text{L},H_\text{em}^\text{det})
}\\
&= \frac{
p(d_\text{gw}|\beta,z_s,d_\text{em}^i,H^i,H_\text{gw}^\text{L})
p(\beta,z_s|d_\text{em}^i,H^i,H_\text{gw}^\text{L})
}{M\cdot
p(d_\text{gw}|d_\text{em},H_\text{gw}^\text{L},H_\text{em}^\text{det})
}\,.
\end{split}
\label{eq:localization_posterior}
\end{equation}

We assign equal prior probabilities to the candidate clusters, $\pi_i^\text{det} \equiv
P(H^i|d_\text{em},H_\text{gw}^\text{L},H_\text{em}^\text{det})=1/M$. Since the
evidence
$p(d_\text{gw}|d_\text{em},H_\text{gw}^\text{L},H_\text{em}^\text{det})$ is
independent of the positional parameters and cluster candidates, the
denominator of posterior can be treated as a constant in this study and does
not affect the localization results.

The $z_s$ prior is adopted to be uniform
for $z_l < z_s \leq5$ and uniform spatial prior is assumed within the cluster model.

For a fixed candidate $i$, we define the localization log-likelihood as:
\begin{equation}
\log L_i(\beta,z_s)\equiv \log p(d_\text{gw}|\beta,z_s,d_\text{em}^i,H^i,H_\text{gw}^\text{L}).
\end{equation}

Assuming Gaussian likelihood, we can express the posterior as simple
log-likelihood (see Appendix~\ref{app:local}):
\begin{equation}
\label{eq:chi-sq}
\begin{split}
  \log L_i(\beta,z_s)
    &= -\sum_{j=1}^{N}\frac{(\DLeff^{\text{obs},j} - \DLeff^{j})^2}{2\sigma_{\DLeff}^2} \\
    &\quad - \sum_{k=1}^{N-1}\frac{(\Delta \tau_{k,\text{obs}} - \Delta \tau_{k,\text{H}})^2}{2\sigma_{\Delta \tau_k}^2},
\end{split}
\end{equation}
where $N$ is the number of images; the hypothesis $\text{H}=(i, \beta, z_s )$;
$\Delta \tau_{k,\text{obs}}$ is the observed $\Delta \tau_k$; $\Delta
\tau_{k,\text{H}}$ is the $\Delta \tau_k$ predicted under the hypothesis. The uncertainties
of $\vDLeff$ and $\Delta \tau$ are denoted as
$\sigma_{\DLeff}$\footnote{We set the luminosity distance uncertainty
$\sigma_{\DLeff}$ to 10 percent to reflect the uncertainty from each
image magnification. This discrepancy is possible due to expected small effects
of stellar microlensing on lensed GWs~\citep{HannukselaOttoA2020Lmbh}.} and
$\sigma_{\Delta \tau}$ respectively. A detailed explanation of these
uncertainties are shown in Appendix\ref{app:maps} and~\ref{app:local}.

We first employ the global optimization method, the differential evolution (DE)
algorithm~\citep{DE}, to obtain an initial estimate of $\text{H}$ with relatively high likelihood,
preventing the sampling procedure from trapping in local maxima. We then perform MCMC by the Python package $emcee$~\citep{emcee_theory,emcee} to explore the posterior distribution and determine the credible intervals of $\text{H}$.

To evaluate the determination of the correct galaxy cluster lens, we compare each candidate with the union
of the other candidates in the catalog. The posterior odds assuming true cluster in the catalog and using the evidence of each candidate is:
\begin{equation}
O_{\rm det}^{i}
\equiv
\frac{P(H^i\mid d,H_{\rm gw}^{\rm L},H_{\rm em}^{\rm det})}
{P(H^{\rm other}\mid d,H_{\rm gw}^{\rm L},H_{\rm em}^{\rm det})}=
\frac{\pi_i^{\rm det}Z_i^{\rm det}}
{\sum_{j\ne i}\pi_j^{\rm det}Z_j^{\rm det}}.
\label{eq:main_full_odds}
\end{equation}
where $d=(d_{\rm gw},d_{\rm em})$, $Z_k^\text{det}
\equiv
p(d_\text{gw}|d_\text{em},H^k,H_\text{gw}^\text{L})$ and $H^\text{other}$ is hypothesis that a lens other than cluster $i$ is the lens.
The derivation is given in
Appendix~\ref{app:posterior_odds_derivation}. The appendix further approximate the odds by maximum likelihood estimators:
\begin{equation}
O_\text{det}^i =
\frac{
Z_i^\text{det}
}{
\sum_{j\neq i}Z_j^\text{det}
}\approx \widetilde{O}_\text{det}^i=\frac{
\exp(\log L_i^*)
}{
\sum_{j\neq i}\exp(\log L_j^*)
}\,,
\label{eq:main_Odet_evidence}
\end{equation}
where $\log L_k^*$ is the best-fit log-likelihood under candidate k:
\begin{equation}
\log L_k^*
=\max_{\beta,z_s}\log L_k(\beta,z_s),
\end{equation}
and we can neglect the Occam factors if the detected candidates have comparable
prior volumes and comparable posterior widths. We thus use $\widetilde{O}_\text{det}^i$ to estimate the confidence of lens
identification. We determine whether a cluster lens is identified by a threshold. The integrated-evidence calculation used for model marginalization is described separately below.

\subsection{Model marginalization}
\label{ssec:model_marginalization}

A cluster can have several published reconstructions that predict
different lensing observables. We study this by marginalizing the evidence of each model. Let $m$ denote the index of one of the available reconstructions of cluster $i$, $\Theta=(\beta_x,\beta_y,z_s)$ and $L_{im}(\Theta)$, $\pi_{im}(\Theta)$ be the normalized likelihood and source prior for
that reconstruction. Using the normalized
likelihood of Appendix~\ref{app:local}, the individual and
model-marginalized evidences are
\begin{equation}
\begin{split}
Z_{im}&=\int L_{im}(\Theta)\pi_{im}(\Theta)\,d\Theta,\\
Z_i^{G}&=\sum_{m}w_{im}Z_{im},
\qquad
\sum_{m}w_{im}=1,
\end{split}
\label{eq:D_model_evidence}
\end{equation}
where $w_{im}$ is the prior model probability and $G$ superscript denotes the marginalized evidence. The adopted prior weights are again uniform over each model.

The cluster-identification odds become
\begin{equation}
O_{{\rm det},{\rm G}}^{i}
=
\frac{\pi_i^{\rm det}Z_i^{\rm G}}
{\sum_{j\ne i}\pi_j^{\rm det}Z_j^{\rm G}}.
\label{eq:D_identification_odds}
\end{equation}
Conditional on cluster $i$, the source posterior is
\begin{equation}
p(\Theta\mid d,H^i)
=
\sum_{m\in\mathcal{M}_i}
P(m\mid d,H^i)\,p(\Theta\mid d,H^i,m),
\label{eq:model_mixture}
\end{equation}
with posterior model weights
\begin{equation}
P(m\mid d,H^i)
=\frac{w_{im}Z_{im}}{Z_i^{\rm MA}}.
\label{eq:posterior_model_weights}
\end{equation}
The evidences of each reconstructions are evaluated using
the same DE and MCMC stages as in Sec.~\ref{ssec:evaluation}. The reconstruction of simulation is retained in the set of reconstructions.

\subsection{Reconstruction-mismatch test and evaluation}
\label{ssec:mismatch_design}

For each event simulated with reconstruction $A$, we attempt the localization with each alternative reconstruction $B\ne A$ of the same cluster.
Here, we perform the tests with four-image events due to the reason described in Sec.~\ref{ssec:simulation}.
We use three parametric models in HSTFF which are CATS, Keeton and Sharon for the tests.
Five events for each cluster for each family of models are localized and we require the predicted image count to match the simulated event as a pilot test.
We record whether each attempt can obtain a valid initial guess on a cluster and the posterior if the correct cluster lens is identified.
We then assess the discrepancies between the injected and localized source position and the coverage of the 90\% credible intervals.

\section{Results}
\label{sec:result}

\begin{figure*}
    \includegraphics[width=\textwidth]{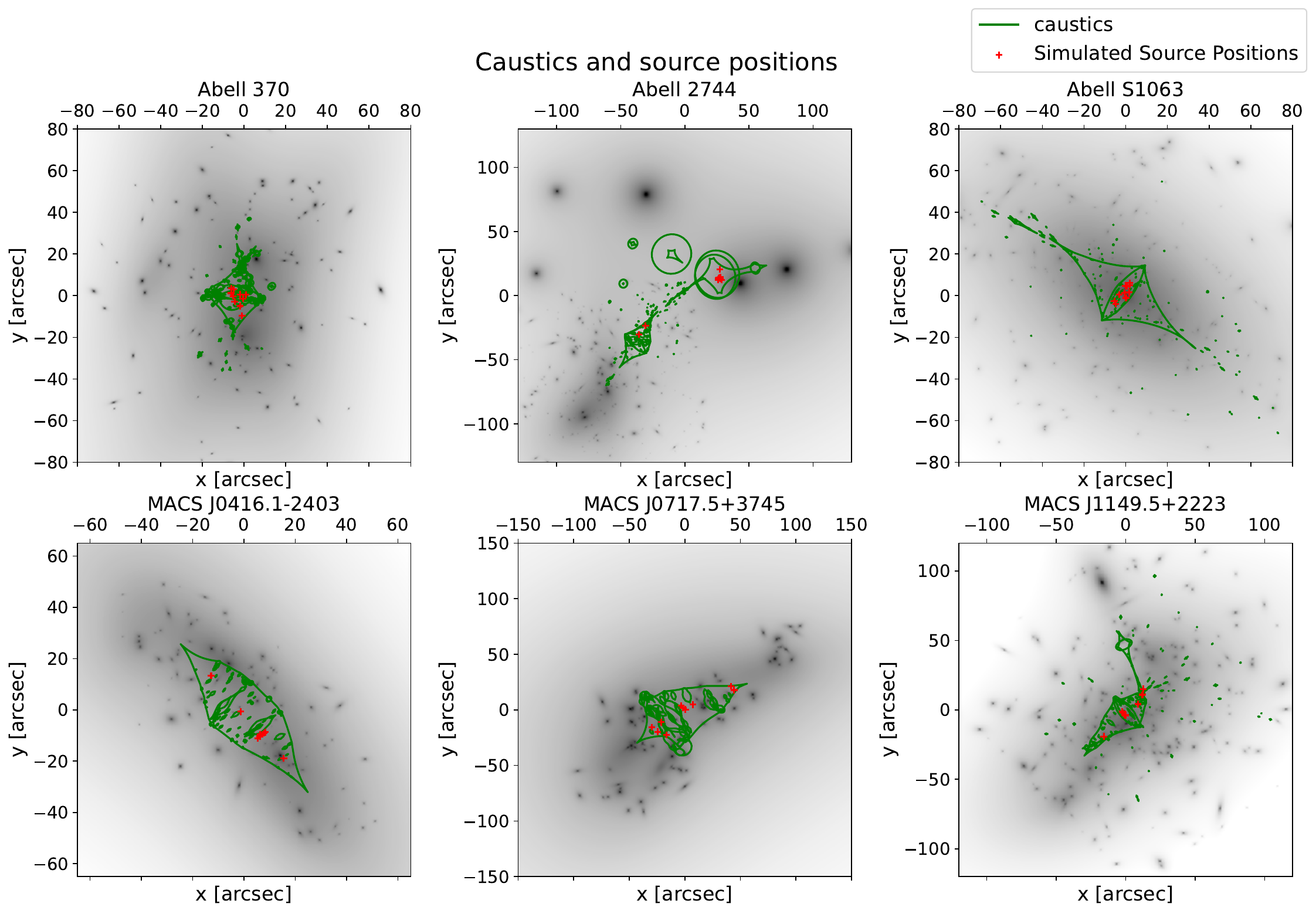}
    \caption{
    	Caustics of the 6 galaxy clusters used in this study with simulated source positions, and the surface mass density. 
	The green curves indicate the cluster caustics, red pluses represent simulated source positions in the source plane. 
	The gray shading represents surface mass density in the image plane for reference. 
	The sources are located inside the inner caustic, often in regions associated with complex critical curves, which permit the formation of multiple lensed images.
	}
    \label{fig:6clusters}
\end{figure*}

The 60 source positions for the idealized analysis and
caustics of the clusters are shown in Fig.~\ref{fig:6clusters}.

We first show the results of matched-model cluster identification (Sec.~\ref{ssec:lens_identification}), followed by the source localization results with $z_s$ sampled (Sec.~\ref{ssec:lens_localisation_within_cluster}) and the results with $z_s$ fixed to its injected value (Sec.~\ref{ssec:lens_localisation_within_cluster_known_z}). In Sec.~\ref{ssec:correlation_of_bayes_factor}, we study the possible correlations between lens identification and properties of both the events and lenses.
Sec.~\ref{ssec:marginalized_results} and Sec.~\ref{ssec:mismatch_results} present the model-marginalized and reconstruction-mismatch results.

\subsection{Identification of Galaxy Cluster Lens}
\label{ssec:lens_identification}
\begin{figure}
    \includegraphics[width=\columnwidth]{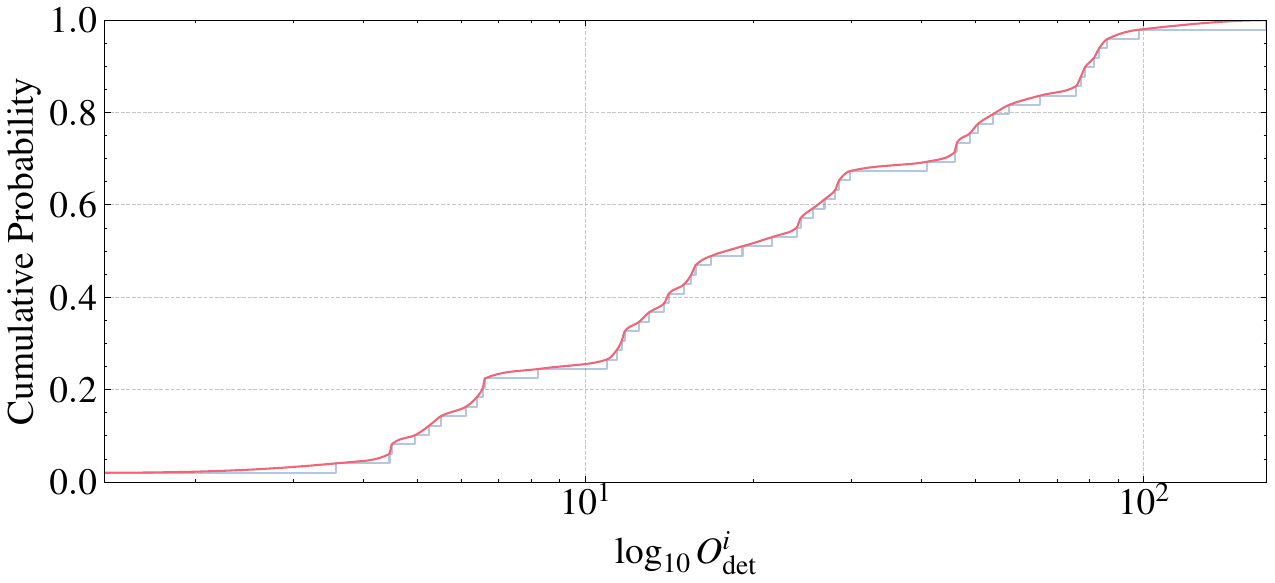}
    \includegraphics[width=\columnwidth]{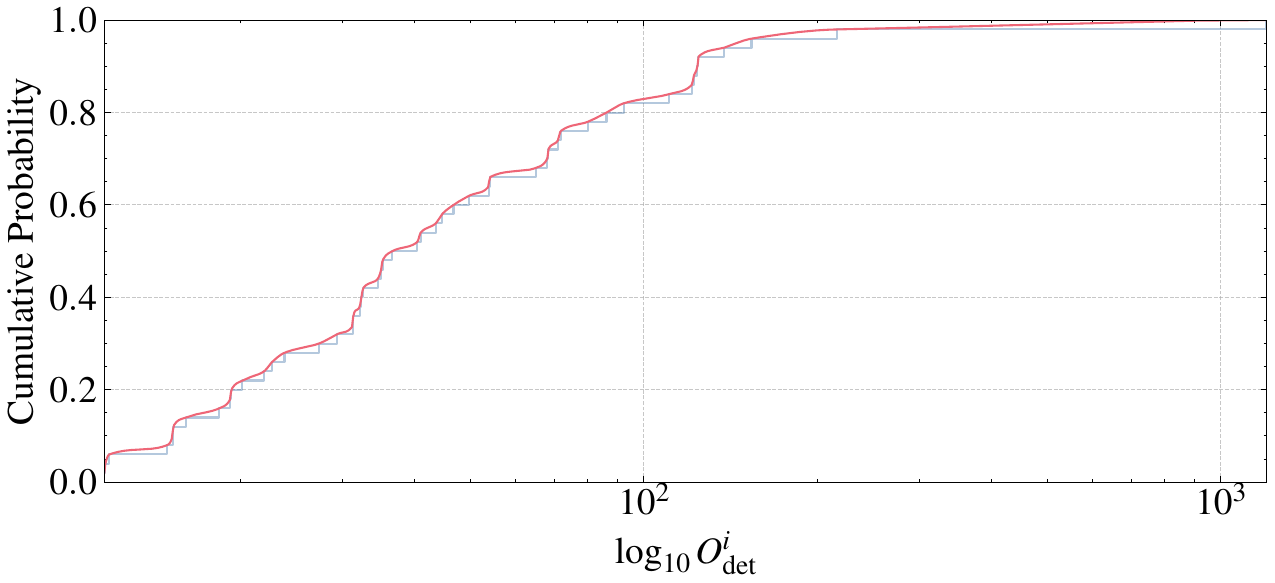}
    \caption{
	Cumulative frequency distribution of posterior odds of lens identification of events with successful localization, with $z_s$ sampling (upper panel) and without $z_s$ sampling (lower panel) in double log-scale for clarity. 
	The posterior odds are $\widetilde{O}_\text{det}^i$ between most (i) and other hypotheses that the corresponding cluster is the lens. 
	For both distributions, $\log_{10} \widetilde{O}_\text{det}^i$ are all > 3, and the $10^{\mathrm{th}}$ percentiles are larger than 5 in both sampling cases, indicating strong confidence of the cluster identification.}
    \label{fig:odds_distribution}
\end{figure}

We determine the successful identification of the galaxy cluster lens by the approximation of posterior odds defined in the method section. We show the identification confidence of galaxy cluster lens of the simulated events with and without sampling $z_s$ by their posterior odds displaying in Fig.~\ref{fig:odds_distribution}.

We find that $88.3\%$ of the simulated events pass the threshold of $\log_{10}\widetilde{O}_\text{det}^i \ge3$.
The failed events are due to the computational failure of DE to find reasonable initial guesses, while one event has wrong lens identification, making the successful events slightly less than $90\%$ of simulated events.
For the simulations with $z_s$ sampled, the posterior odds values are smaller than simulations with known $z_s$, which is expected due to larger parameter space volume when sampling in higher dimensions.
For both types of simulations, successful localized events all have $\log_{10} \widetilde{O}_\text{det}^i > 3$.
Also, $\log_{10} \widetilde{O}_\text{det}^i>$ $5$ and $10$ for around 10th percentile of the localized event with and without $z_s$ sampled respectively, denoted by the shaded region in the figures.
This indicates strong confidence of determining the correct cluster lens for most of the simulated events.

\subsection{Localization Results with unknown redshift}
\label{ssec:lens_localisation_within_cluster}

We jointly infer $(\beta_x,\beta_y,z_s)$ where the source redshift is not known before identifying the host galaxy of the GW source. The lens identification and localization performance is summarized in Table~\ref{tab:general-result_xyz}.

\begin{table}
  \centering
  \caption{
    Summary of the localization results with $z_s$ sampled. The number of
    simulated events and the number of events with successful lens identification
    and localization are shown. The corresponding median widths of the 90\%
    intervals for each sampled parameter ($\beta_x,\beta_y,z_s$) are also shown.
  }
  \label{tab:general-result_xyz}

  \small
  \setlength{\tabcolsep}{3pt}
  \begin{tabular*}{\columnwidth}{@{\extracolsep{\fill}} *{5}{c} @{}}
    \toprule
    No.\ events & Identification
    & $\overline{\delta \beta_x}$
    & $\overline{\delta \beta_y}$
    & $\overline{\delta z_s}$ \\
    \midrule
    60 & 53 & $0.33''$ & $0.33''$ & $0.27$ \\
    \bottomrule
  \end{tabular*}
\end{table}

The table demonstrates that the localization algorithm achieves high precision. We identify the correct cluster in 53 of the 60 events. Four of these 53 events fail the numerical acceptance criterion for the MCMC fit\footnote{We determine the numerical issue for an event by using $L_k^*$ as a numerical estimator and set the maximum value of $L_k^*$ in order to claim convergence.}, leaving 49 events for source localization. 
For these 49 events, the median widths of the marginal 90\% intervals are $0.33''$ in each source coordinate and $0.27$ in redshift.
Thus, the position precision are in sub-arcsec precision.
Besides, the table~\ref{tab:general-result_xyz} shows the median uncertainty of $\overline{\delta z_s}$, which is relatively large comparing to the injected value of $z_s$.
Although this constraint is less precise than a spectroscopic redshift measurement, it provides useful information
when searching for the host galaxy, especially when the inferred redshift is combined with the source-plane localization to narrow the possible host galaxy candidates.

\begin{figure}
    \includegraphics[width=\columnwidth]{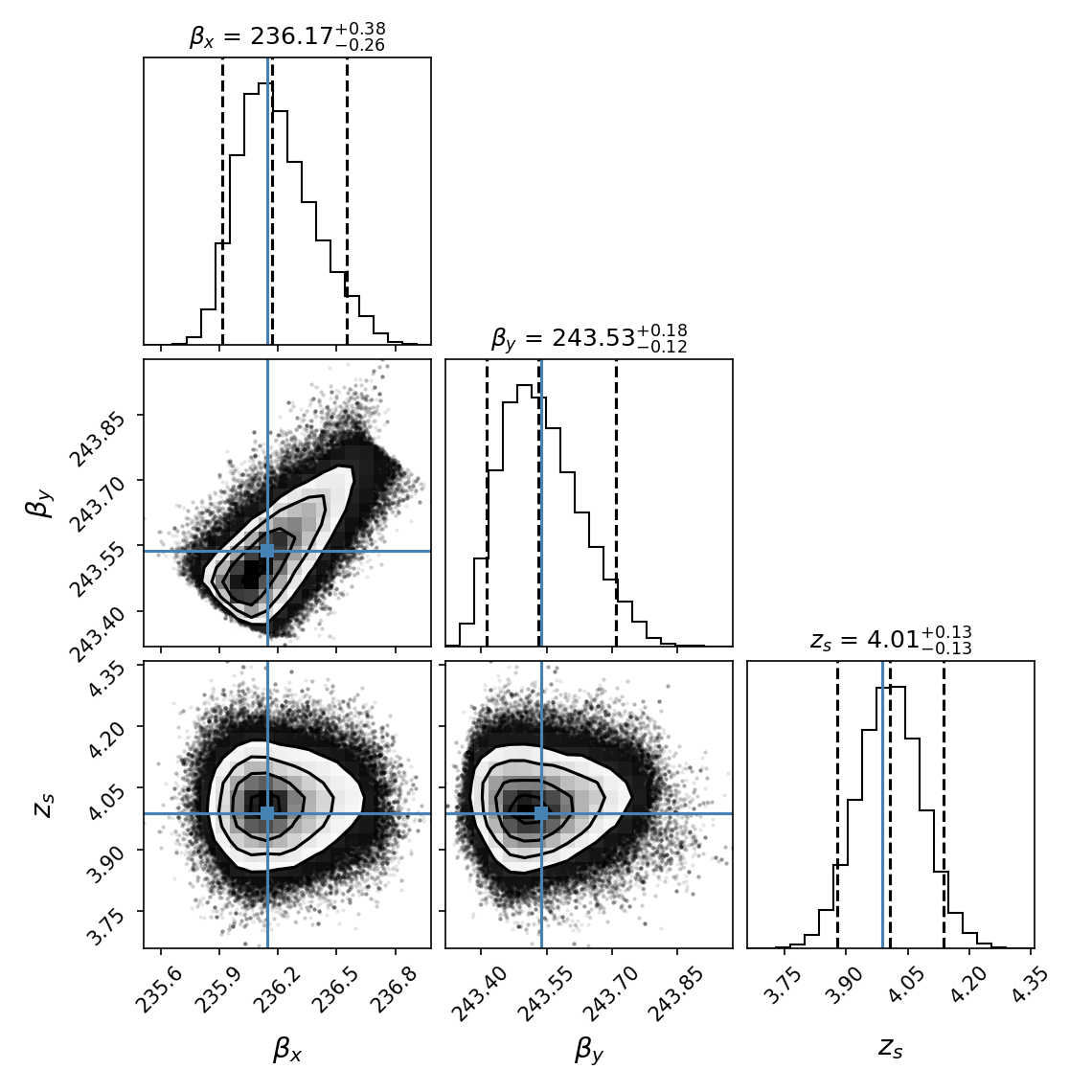}
    \caption{An example of corner plot of a simulated event with MACS
	    J0717.5+3745 as lens and $z_s$ sampling. The blue cross is the true
	    value of the source position, and the dotted lines mark the $90\%$
	    credible intervals. The true parameter values are within the
	    credible intervals of the localization, with uncertainties of
	    ($\beta_x$, $\beta_y$) below $\pm 0.4\, \mathrm{arcsec} $, and $\pm
	    0.13$ for $z_s$. The sources are localized to sub-$\mathrm{arcsec}$ precision, while $z_s$ is less constrained.
    }
    \label{fig:corner_xyz}
\end{figure}

Fig.~\ref{fig:corner_xyz} shows the corner plot of one event behind MACS J0717.5+3745 as an example. The sampled marginal posteriors are single-peaked, and the injected value lie within 90\% interval for each parameter.

\begin{figure}
    \centering
    \includegraphics[width=\columnwidth]{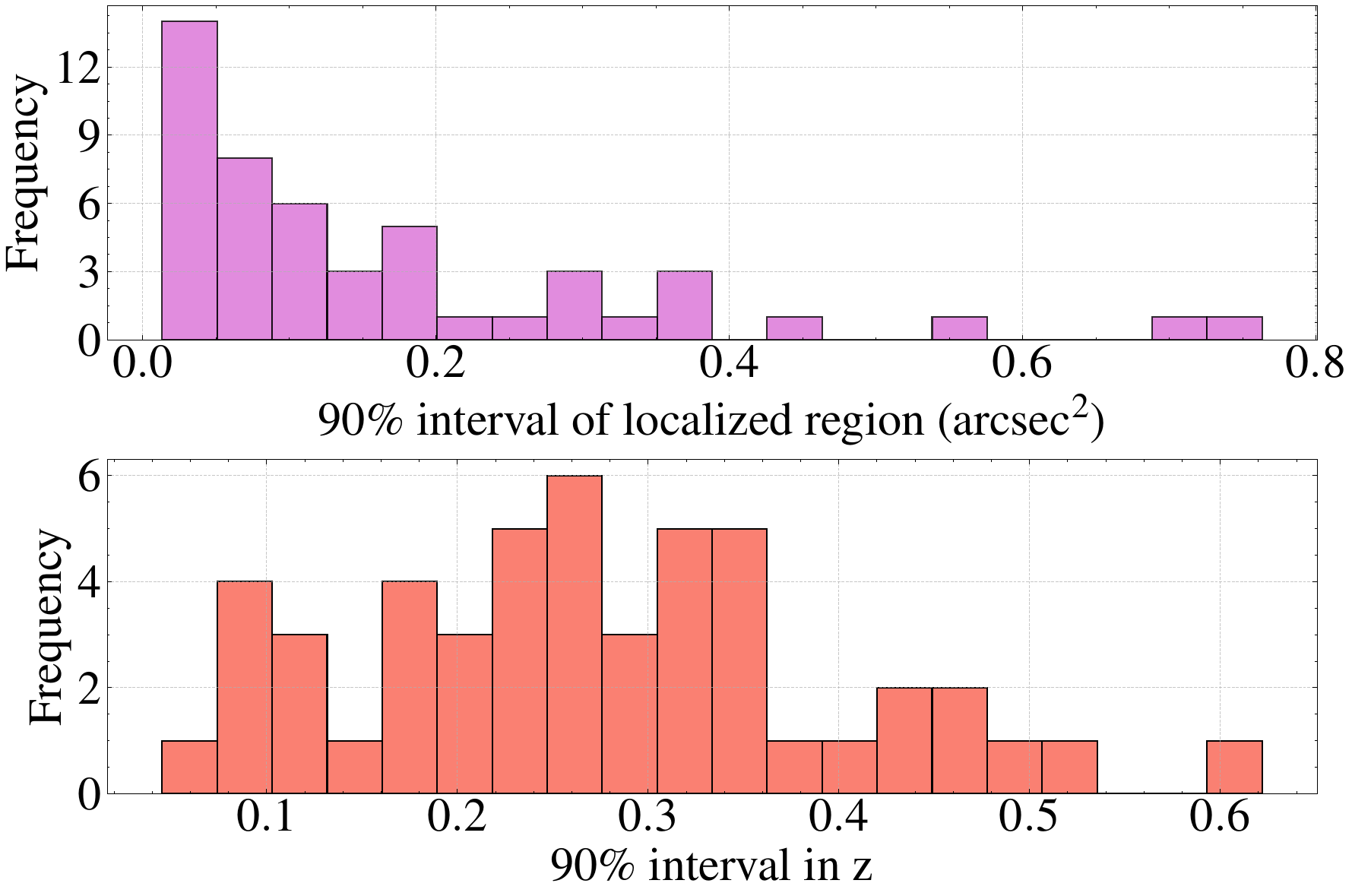}
    \caption{Distribution of the localization-scale proxy $A_{\rm loc}=\delta\beta_x\,\delta\beta_y$ and of the marginal 90\% interval width of $z_s$.
    	The upper panel shows the product of the marginal 90\% widths of $\beta_x$ and $\beta_y$ (not a joint 90\% credible area).
	Most events have $A_{\rm loc}$ below 0.4 $\mathrm{arcsec}^2$.
	The lower panel shows the corresponding $z_s$ interval widths from lensed GW data alone.}
    \label{fig:uncertainty_distribution}
\end{figure}

For the localization uncertainties, we also show the distribution of the localized region and $z_s$ uncertainties for each simulated event in Fig.~\ref{fig:uncertainty_distribution}.
We define $A_{\rm loc}=\delta\beta_x\,\delta\beta_y$ as the product of the marginal 90\% interval widths of $\beta_x$ and $\beta_y$; this is a convenient axis-aligned scale and is not a joint 90\% credible area in the source plane.
The histogram bars show $A_{\rm loc}$ and the marginal 90\% interval widths of $z_s$ for each event, excluding failed localizations.
Most events have $A_{\rm loc}$ below 0.4 $\mathrm{arcsec}^2$, and most of the 90\% interval widths of $z_s$ span from around $0.1$ to $0.5$, which is large considering the values of $z_s$.
Therefore, while the source-plane constraints reach $\mathrm{arcsec}^2$ scales in this proxy, the $z_s$ results carry larger uncertainties and serve as rough estimations of their true values.

In summary, we can identify the galaxy cluster lens of the lensed GW event, constrain the source position with sub-$\mathrm{arcsec}^2$ precision and estimate $z_s$ without a prior identification of its host galaxy. The inferred source position and redshift can then be compared with EM reconstructions of galaxy members in the cluster to narrow down the possible hosts.

\subsection{Localization results with known redshift}
\label{ssec:lens_localisation_within_cluster_known_z}

We next investigate a follow-up scenario where $z_s$ has been measured independently by EM observations and infer only
$(\beta_x,\beta_y)$.
To quantify the potential improvement on source localization and lens identification, we fix $z_s$ to its injected value and infer only the source coordinates $(\beta_x,\beta_y)$. Table~\ref{tab:general-result_xy} summarizes the results.

\begin{table}
  \centering
  \caption{
      Summary of the localization results with known $z_s$. Same types of data
      are shown as in Table~\ref{tab:general-result_xyz}. The median widths of
      90\% intervals are approximately $0.26\, \mathrm{arcsec}$ in $\beta_x$
      and $\beta_y$, indicating higher precision in sky localization than the
      case with $z_s$ sampling.
    }
    \label{tab:general-result_xy}
    \small
  \setlength{\tabcolsep}{3pt}
  \begin{tabular*}{\columnwidth}{@{\extracolsep{\fill}} *{4}{c} @{}}
    \toprule
    No.\ events & Identification
    & $\overline{\delta \beta_x}$
    & $\overline{\delta \beta_y}$ \\
    \midrule
    60 & 54 & $0.26''$ & $0.26''$ \\
    \bottomrule
  \end{tabular*}
\end{table}

Again, only localization results of 50 successful events are shown due to MCMC numerical convergence issues.
The success rates of both lens identification and localization are slightly higher than that in the unknown-$z_s$ analysis.
Compared with the unknown-redshift scenario, the median marginal 90\% interval width is reduced to $0.26''$ in each source coordinate.
Fixing the source redshift removes one degree of freedom and restricts the range of source configurations compatible with the observed lensing properties.
The posterior odds of lens identification also show substantial increase.
These improvements demonstrate the potential of EM counterpart redshift measurement to further constrain the source position.

\begin{figure}
    \includegraphics[width=\columnwidth]{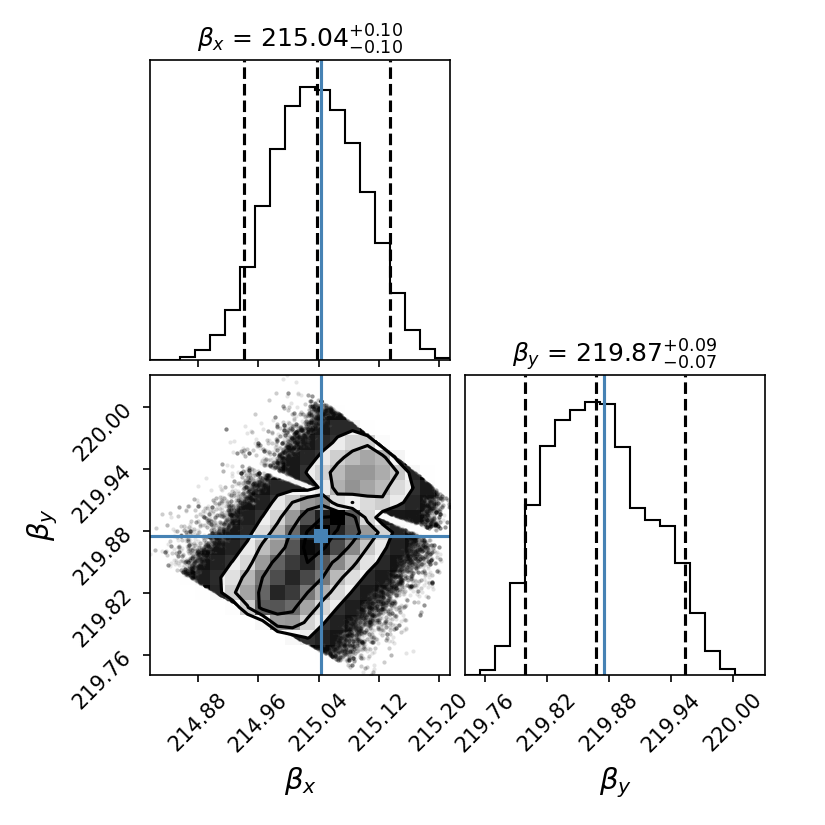}
    \caption{
	    An example of corner plot of a simulated event with MACS
	    J0717.5+3745 as lens and known $z_s$. The blue cross is the true
	    value of the source position, and the dotted lines mark the $90\%$
	    credible intervals. The true parameter values are within the
	    credible intervals of the localization, with uncertainties of
	    ($\beta_x$, $\beta_y$) below $\pm 0.1\, \mathrm{arcsec} $. The
	    source position is more tightly constrained than in the case where
	    $z_s$ is sampled, showing even higher localization precision.
    }
    \label{fig:corner_xy}
\end{figure}

A corner plot for a typical localized event is shown in
Fig.~\ref{fig:corner_xy} as an example similar to the previous section.
The posteriors of all parameters are single-peaked, indicating no distinct degeneracies in this sampling, which is also the case for most of the performed localizations sampling only $\vec{\beta}$.

\begin{figure}
    \centering
    \includegraphics[width=\columnwidth]{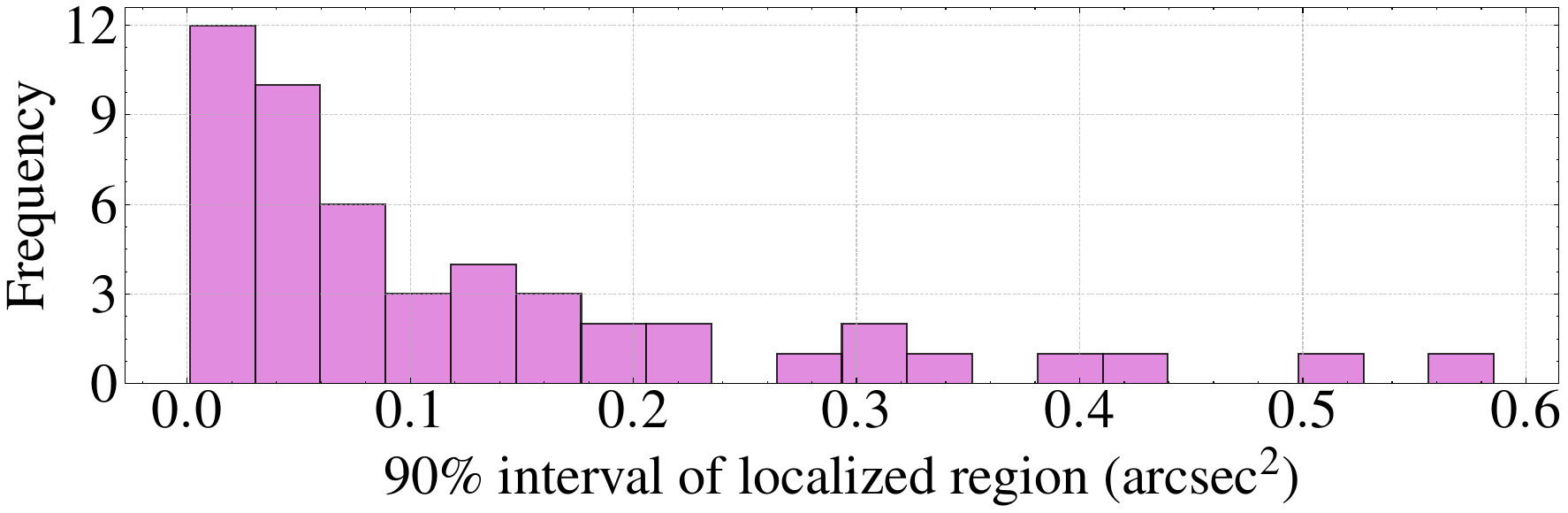}
    \caption{
	   Distribution of the localization-scale proxy
	    when $z_s$ is fixed.
	    As in Fig.~\ref{fig:uncertainty_distribution}, this is the product of
	    marginal 90\% widths rather than a joint 90\% credible area.
	    Most events have $A_{\rm loc}$ below $0.2$--$0.3\,\mathrm{arcsec}^2$.
	}
    \label{fig:uncertainty_distribution_xy}
\end{figure}

We again report the localization-scale proxy
$A_{\rm loc}=\delta\beta_x\,\delta\beta_y$ constructed from the marginal 90\% widths of the source-position parameters, and display the
distribution in Fig.~\ref{fig:uncertainty_distribution_xy}.
Most events have $A_{\rm loc}$ below $0.2\,\mathrm{arcsec}^2$, smaller than in the free-$z_s$ analysis because one degree of freedom is removed.

An independent redshift constraint could therefore improve positional precision within this framework.
We emphasize that fixing $z_s$ to its true value represents an idealized scenario in which $z_s$ is independently determined. In practice, an EM redshift measurement would be associated with the GW observations.
It could introduce finite $z_s$ range to help improving the localization which is more realistic than locating the exact $z_s$.

\subsection{Correlation of Lens identification Confidence and Event Properties}
\label{ssec:correlation_of_bayes_factor}

We analyze the possible correlations between the lens identification confidence
represented by $\log_{10}\widetilde{O}_\text{det}^i$ and the properties of
events and their respective lens models here. For the properties of events, we
inspect the relationship between $\log_{10}\widetilde{O}_\text{det}^i$ and
maximum relative time delay among image pairs and the largest image
magnification of each simulated event. There is no clear dependence of
posterior odds with $\Delta\tau$ and $|\mu|$, suggesting single summary
statistics of the observables may not predict identification confidence. $z_s$
also has no clear correlation with $\log_{10}\widetilde{O}_\text{det}^i$ when
it is also sampled.

We then analyze the possible correlations between the cluster lens model
properties and the posterior odds. One of the important properties is the
relative time delay uncertainty of image positions $\sigma_{\Delta \tau}$,
which can represent the total uncertainties in $\Delta \tau_k$ estimations. We
show the spatial distribution of it in Appendix~\ref{app:maps}. From the plot,
the region with higher $\sigma_{\Delta \tau}$ are mostly in the center of the
clusters as those are the regions with more complex lensing structures.

\begin{figure}
    \centering
    \includegraphics[width=\columnwidth]{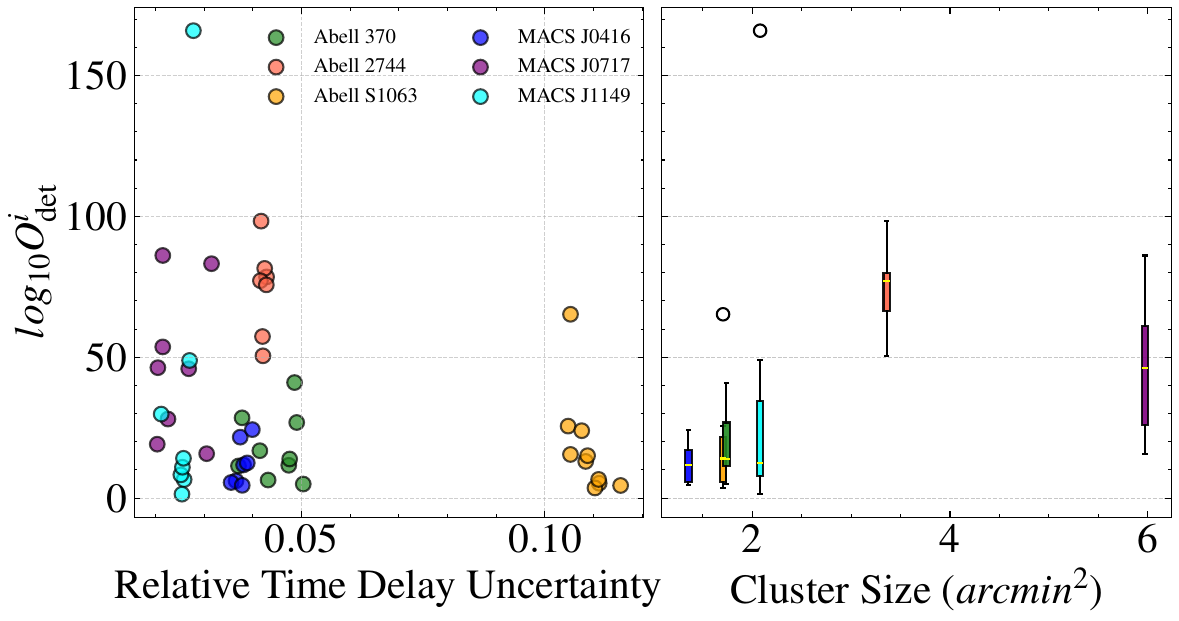}
    \includegraphics[width=\columnwidth]{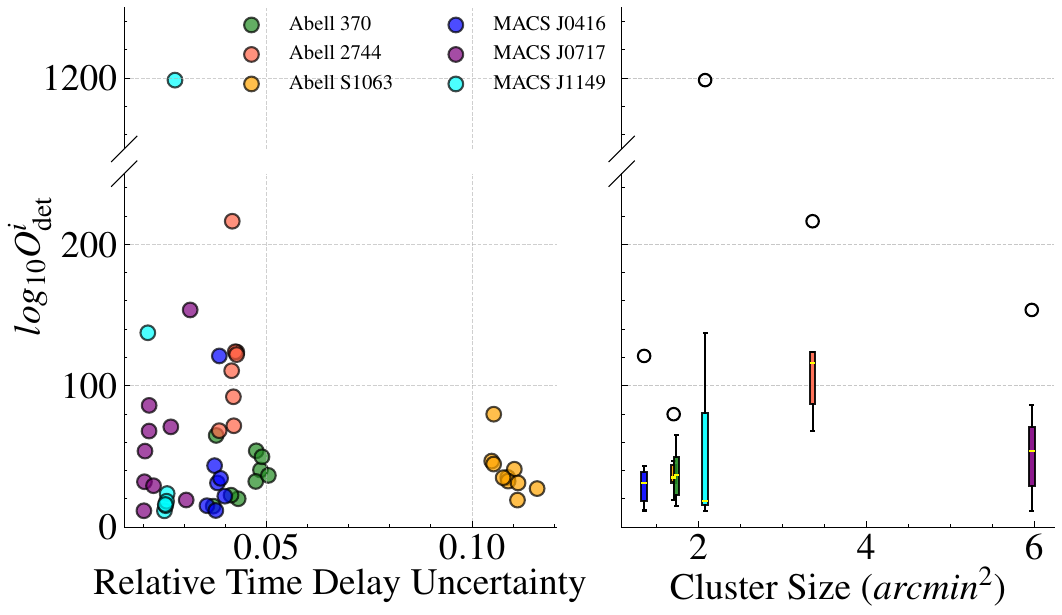}

    \caption{
    Correlation plot of $\log_{10}\widetilde{O}_\text{det}^i$  with the cluster
    properties: the largest $\sigma_{\Delta\tau_k}$ and the size of the
    cluster. The upper panel shows the plots of localization results with $z_s$
    sampling, and lower panel shows the results with known $z_s$.
    There is positive correlation (Spearman correlation 0.3-0.5 with p<0.05) between apparent cluster size and lens-identification confidence.
    }
    \label{fig:O_vs_dt_size_z_xyz}
\end{figure}

We hence display the correlation plots between the lens model properties and
$\log_{10}\widetilde{O}_\text{det}^i$ in Fig.~\ref{fig:O_vs_dt_size_z_xyz}.
Shown in the figure, $\log_{10}\widetilde{O}_\text{det}^i$ has no clear
correlations with the largest $\sigma_{\Delta_t}$ among the images for each
event. However, for the apparent cluster size\footnote{we estimate the apparent
cluster sizes by their critical curve area}, both lens identifications with or
without $z_s$ sampling show clear correlation ($p<0.05$) with the posterior
odds, with Spearman correlation coefficient 0.57 (sampling with $z_s$) and 0.3
(sampling without $z_s$). 
Therefore, there is a moderately strong correlation
between the apparent size of a cluster lens and
its identification confidence.

\subsection{Model-marginalized identification and localization}
\label{ssec:marginalized_results}

\begin{figure}
    \centering
    \includegraphics[width=\columnwidth]{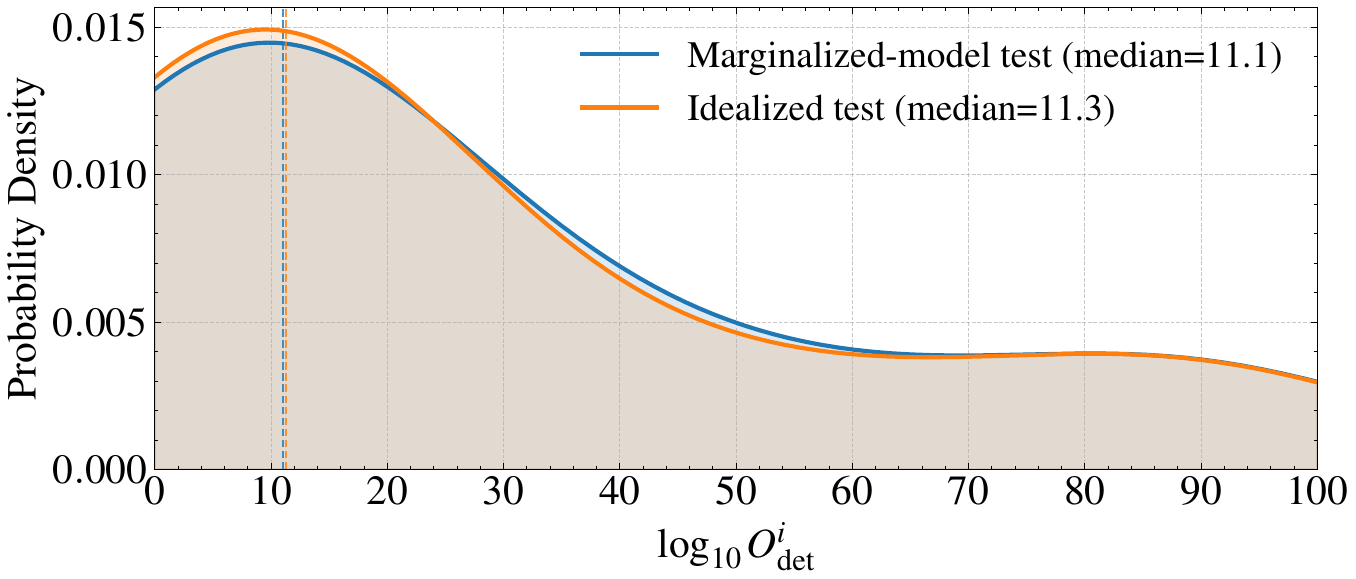}

    \caption{
    Probability density of $\log_{10} \widetilde{O}_\text{det}^i$ of select events in marginalized-model test, comparing with the values in idealized test.
	The median of $\log_{10} \widetilde{O}_\text{det}^i$ of marginalized-model test is slightly lower than that of the idealized test, suggesting lower confidence in multiple-model situation.
    }
    \label{fig:odds_marginalize}
\end{figure}

We perform tests of lens identification and localization where multiple models of the same cluster is marginalized to account for the model discrepancies.
We select 18 five-image events that were successfully localized in the matched-reconstruction analysis. For each event, we average over the available reconstructions of each cluster, including the reconstruction used to generate the event. All 18 events retain the correct unique cluster identification and the source localization. The posterior odds of lens identification are shown in Fig.~\ref{fig:odds_marginalize}.

Note that the median of log posterior odds ($11.1$) is slightly lower than that of the idealized test ($11.3$).
The possible reason is due to the inclusion of evidence of wrong model in the calculation of $Z_i^{G}$, diluting the overall confidence of lens identification of one cluster lens.
The localization results are the same as in the idealized test due to the correct cluster lens and models included in this test.
This test shows that the lens identification and localization remains possible with several available models when the set includes an adequate reconstruction, while the confidence might be slightly lowered.

\subsection{Reconstruction-mismatch results}
\label{ssec:mismatch_results}

\begin{figure}[h]
    \centering
    \includegraphics[width=\columnwidth]{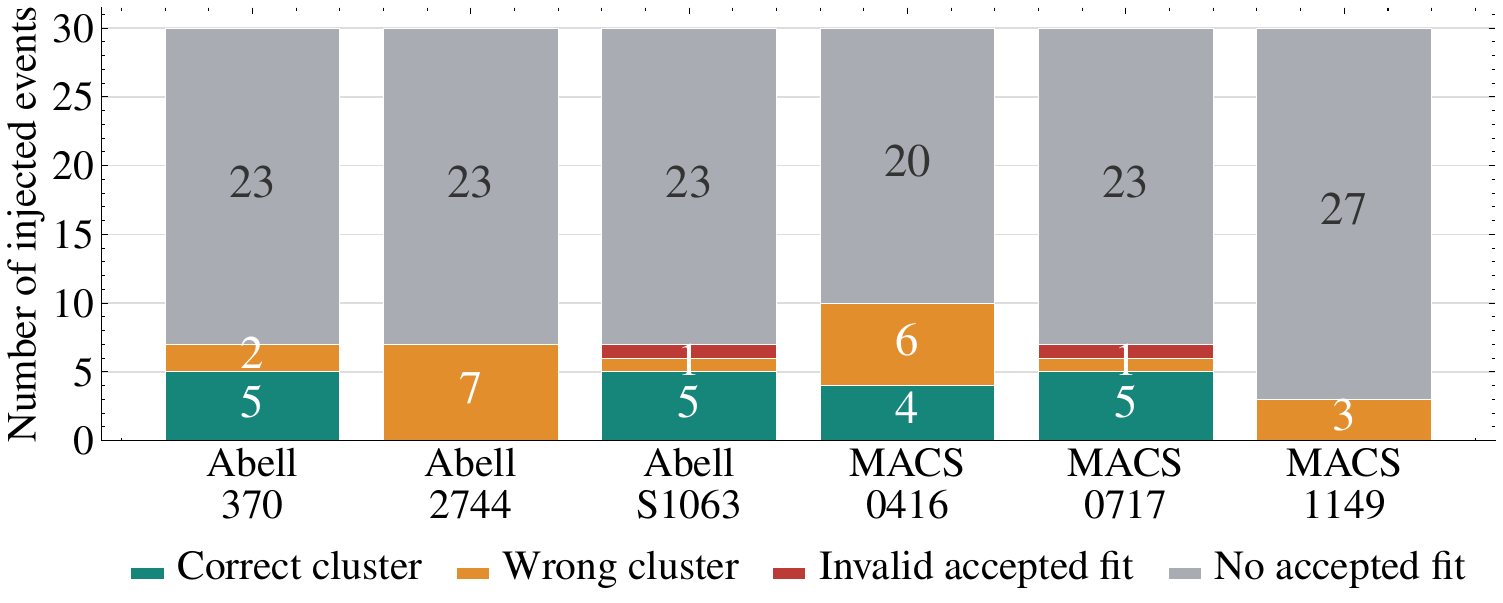}
    \includegraphics[width=\columnwidth]{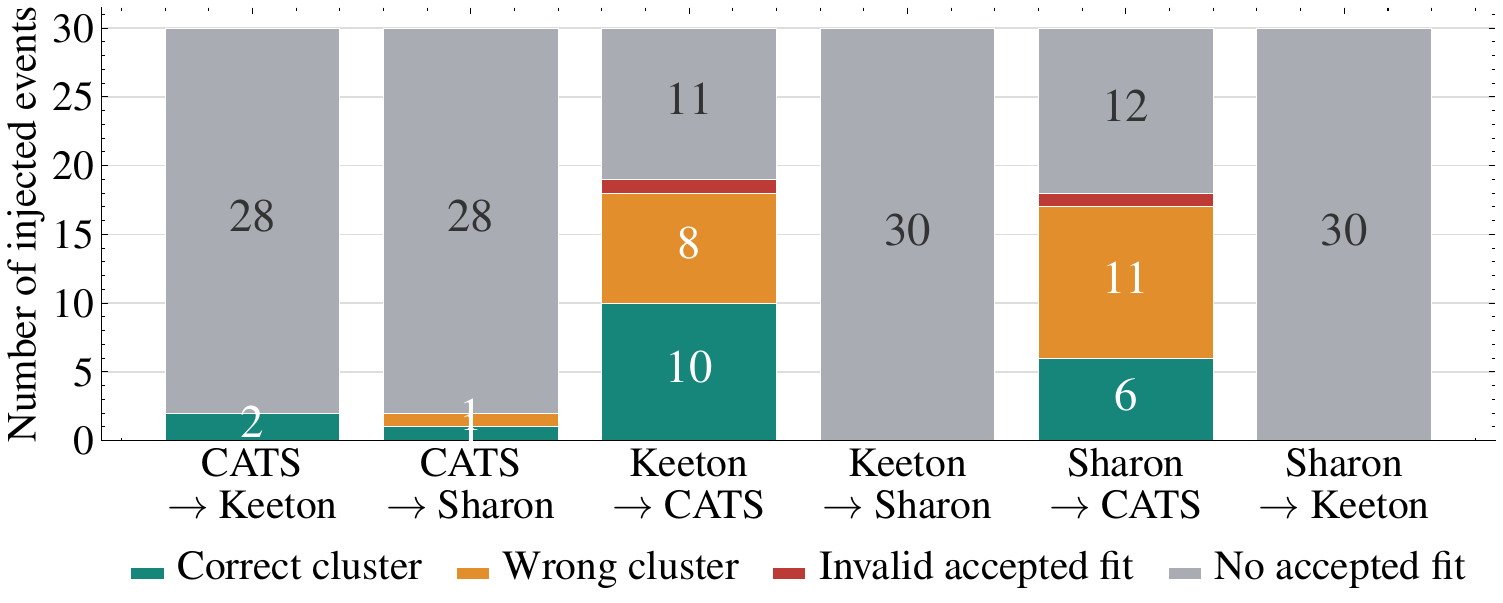}

    \caption{
    Distribution of successful and failed events in lens identification when there are mismatches of reconstruction used in simulation and localization.
    Distribution of such events per cluster (upper panel) and per model pair (lower panel) is shown.
    Around 90\% of localization attempts fail to identify the correct cluster lens.
    Therefore, the localization generally fails in the reconstruction-mismatch setting.
    }
    \label{fig:proportion_mismatch}
\end{figure}

We finally test the localization when all reconstructions does not reliably identify the correct cluster by simulating events with one cluster model and attempt to perform localization by other available models of that same cluster. We report per pair of models and cluster, the fraction of events where the correct cluster is identified. The results are presented in Fig.~\ref{fig:proportion_mismatch}.

As shown in the figure, only 10.5\% of tests can identify the correct cluster lens.
Most attempts failed as no cluster candidate yields an initial fit; the remaining failed events have wrong cluster lens identified or invalid fit after MCMC when an initial fit exists.
Due to the low proportion of successful tests even in lens identification, the successful localization fraction is also low.
Therefore, we conclude that the localization fails in general when there is no adequate reconstruction that describes the cluster lens structure well enough.

\section{Discussion and Conclusion}
\label{sec:conclude}

Gravitational-wave lensing is forecasted to become detectable in the coming years, with current forecasts based on population-I/II stars and the known population of galaxies yielding between one in a few hundred to one in a thousand events being strongly lensed.
Lensing is expected to become commonplace with the third-generation detectors~\cite{Li_2018,PhysRevD.97.023012,2024arXiv240707526P}.
Although galaxy cluster lensing is predicted to occur less frequently than galaxy-scale lensing~\cite{Hilbert2007StrongLensing,Hilbert2008StrongLensing, Smith_2023,Vujeva2025EffectsOf}, the optical depth may be more dominant at higher magnification owing to the long caustic curves in galaxy clusters~\cite{Hilbert2007StrongLensing,Hilbert2008StrongLensing}.
With around $10^5-10^6$ total mergers in the third-generation detectors, galaxy cluster lensing should be observed in the future with third-generation detectors.
The increasing number of expected cluster-lensed GW signals observed in future motivates investigation of cluster lenses as a complementary population of strong gravitational lenses.

In this study, we demonstrate the feasibility of identifying galaxy cluster lenses and precisely localizing the sources of strongly lensed GW events using numerical lens models of galaxy cluster lens candidates.
When the same cluster models are used for simulation and localization, correct cluster lens is identified in 53 out of 60 simulated events and the precision of localization can reach sub-$\mathrm{arcsec^2}$ scale.
When $z_s$ is sampled, the injected values lie in the 90\% credible intervals with narrow width of $0.33"$ in each source coordinate.
Therefore, multi-image GW observations, combined with EM cluster lens reconstructions, may provide precise source-plane localization without requiring prior knowledge of the source redshift.
An independent measurement of the source redshift can further strengthen the lens association and tighten the localization constraints.

The five-image configurations considered in this study are motivated by the high image multiplicities possible in galaxy cluster lenses. 
The lensing rate of different image multiplicities in galaxy cluster
lenses has not been studied extensively. 
Nevertheless, such events are plausible when the GW source
lies near specific regions of a galaxy cluster, as illustrated in
Fig.~\ref{fig:five_image_gw_illustration}.
Such configuration is also observed by EM (\cite{Dahle_2013}).
We adopt five-image systems to explore the additional
information available from high-multiplicity cluster lenses.
Future work can extend this analysis to events with fewer detected images, as the information provided by effective luminosity distances can compensate for the reduced number of measured time delays.

All 18 previously localized events retain the correct cluster identification and source localization when we marginalize over multiple lens reconstructions that include the generating model. However, if the alternative models have negligible posterior weight, the resulting posterior may be nearly same as the posterior of the model used in simulation. The discrepancies between models are also not quantitatively analyzed.
Therefore, the results in this test suggest the feasibility of localization in multi-model scenario instead of verifying it in general.

Lens identifications and localizations are less successful in the four-image test with the generating reconstruction excluded. Most attempts fail during initialization.
Also, this pilot study which adopts four-image events may not be aligned with the situation of five-image events; we also only use parametric models and free-form models are not tested yet.
In realistic lensed GW detections, we are not able to guarantee an adequate lens reconstruction is available.
Therefore, whether the localization framework can be applied with current status of EM lens observations and reconstructions remains unresolved and further studies needed to be done to address the robustness of this localization method.

Observational developments motivate applications of this localization framework beyonf the six HSTFF clusters investigated here.
Euclid's first strong-lensing cluster catalog demonstrates the potential of wide-field imaging to expand the sample of candidate cluster lenses \citep{Euclid2026ClusterCatalogue}.
We test the localization method on six HSTFF clusters, while the localization method in general requires a complete all-sky catalog of cluster lenses.
Future observational missions such as Euclid may hence fulfill the data required for this localization.
More well-observed clusters may also improve the quality of lens reconstruction, which can be beneficial to the localization method.

With the successful localization of GW events lensed by galaxy clusters, and current observations of such clusters, existing and future GW events can be searched retrospectively to investigate the possibility of lensing by galaxy clusters.
If such signals are detected, we can precisely constrain their source positions and estimate the source redshift, which can then locate the host galaxy.
With further studies such as a full cluster-lensed GW detection analysis, the localization framework may enable detailed studies of galaxy cluster lenses, host galaxies and support multi-messenger studies of such lensed GW events.

\begin{acknowledgments}
This work utilizes gravitational lensing models produced by PIs Bradač,
Natarajan \& Kneib (CATS), Merten \& Zitrin, Sharon, Williams, Keeton,
Bernstein and Diego, and the GLAFIC group. This lens modeling was partially
funded by the HST Frontier Fields program conducted by STScI. STScI is operated
by the Association of Universities for Research in Astronomy, Inc. under NASA
contract NAS 5-26555. The lens models were obtained from the Mikulski Archive
for Space Telescopes (MAST). H.Y.C., L.E.U., S.J., S.L., and O.A.H.\ acknowledge support by grants from the Research Grants Council of Hong Kong (Project Nos.\ CUHK~14304622, 14307923, and 14307724), the start-up grant from the Chinese University of Hong Kong, and the Direct Grant for Research from the Research Committee of The Chinese University of Hong Kong.
L.E.U.\ is supported by the Hong Kong PhD Fellowship Scheme (HKPFS) from the Hong Kong Research Grants Council (RGC).

\end{acknowledgments}

\newpage
\appendix

\renewcommand*{\theHequation}{appendix.\Alph{section}.\arabic{equation}}

\onecolumngrid

\section{Posterior odds derivation}
\label{app:posterior_odds_derivation}

In order for the GW to be lensed by galaxy cluster $i$ in our catalog of electromagnetic data, the GW event must be established as a strong-lensing candidate, the lens that produced the GW must then be identified, and the GW must originate from cluster $i$. 
The probability that this occurs 
\begin{equation}
	P(H^i, H_\text{em}^\text{det}, H_\text{gw}^\text{L}| d)\,,
\end{equation}
where $H^i$ is the hypothesis that cluster $i$ lensed the gravitational wave, 
$H_\text{em}^\text{det}$ is the hypothesis that the lens is detected by EM observations, and 
$H_\text{gw}^\text{L}$ is the hypothesis that the gravitational wave is lensed. 
The data $d = \{d_\text{gw}, d_\text{em}\}$ denotes all the available information from both EM and GW observations, including the GW strain data and the EM data of all the galaxy clusters in the catalog.\footnote{That is, $d = \{d_\text{gw}, d_\text{em}^1, d_\text{em}^2, ..., d_\text{em}^N\}$ where $N$ is the number of clusters in the catalog.} 
This scenario should be favored over the alternative, that either 
(a) the gravitational-wave is not lensed, or 
(b) it is lensed but its lens system is not detected in the electromagnetic band, 
or 
(c) it is lensed and its lens system is detected in the electromagnetic band, but it is not cluster $i$ that lensed the gravitational wave, but instead any other cluster $j$ in the catalog. 
The probability of this alternative scenario 
\begin{equation}
	P(H^\text{other}, H_\text{em}^\text{det}, H_\text{gw}^\text{L}| d) + P(H_\text{em}^\text{nondet}, H_\text{gw}^\text{L}| d) + P(H_\text{gw}^\text{U}| d)\,,
\end{equation}
where $H_\text{em}^\text{nondet}$ is the hypothesis that the lens is not detected by EM observations, and $H_\text{gw}^\text{U}$ is the hypothesis that the GW is not lensed, and $H^\text{other}$ is the hypothesis that the lens is detected by EM observations and the GW is lensed, but it is not cluster $i$ that produced the GW, but instead any other cluster $j$ in the catalog: 
\begin{equation}
	P(H^\text{other},H_\text{em}^\text{det},H_\text{gw}^\text{L}| d) = \sum_{j\neq i} P(H^j, H_\text{em}^\text{det}, H_\text{gw}^\text{L}| d)\,.
\end{equation}
The posterior odds in favor of cluster $i$ having produced the GW is then given by the ratio of these two probabilities, 
\begin{equation}
	O^i = \frac{P(H^i, H_\text{em}^\text{det}, H_\text{gw}^\text{L}| d)}{P(H^\text{other}, H_\text{em}^\text{det}, H_\text{gw}^\text{L}| d) + P(H_\text{em}^\text{nondet}, H_\text{gw}^\text{L}| d) + P(H_\text{gw}^\text{U}| d)}\,.
\end{equation}

We then assume that the GW lensing has been confirmed and use the chain rule, so the posterior odds
\begin{equation}
	O^i \approx \frac{P(H^i| d,H_\text{gw}^\text{L},H_\text{em}^\text{det})}{P(H^\text{other}| d,H_\text{gw}^\text{L},H_\text{em}^\text{det}) + R_\text{nondet}}\,, \quad R_\text{nondet}
\equiv
\frac{
P(H_\text{em}^\text{nondet}|d,H_\text{gw}^\text{L})
}{
P(H_\text{em}^\text{det}|d,H_\text{gw}^\text{L})
}\,.
\end{equation}

We consider five-image lensed gravitational-wave signals, which typically require massive cluster lenses~\citep{Menghetti2007,Hilbert2007StrongLensing}. Such massive systems are comparatively rare and are well covered by existing deep and wide surveys at the relevant masses and redshifts~\citep{HSTFF,cats_richard_2014}. Therefore, we treat $R_\text{nondet}$ as negligible for the events considered here.

We then define the detected-candidate odds
\begin{equation}
O_\text{det}^i
\equiv
\frac{
P(H^i|d,H_\text{gw}^\text{L},H_\text{em}^\text{det})
}{
P(H^\text{other}|d,H_\text{gw}^\text{L},H_\text{em}^\text{det})
}=\frac{
\pi_i^\text{det}Z_i^\text{det}
}{
\sum_{j\neq i}\pi_j^\text{det}Z_j^\text{det}
}\, ,
\label{eq:Odet_def}
\end{equation}
where $\pi_k^\text{det}
\equiv
P(H^k|d_\text{em},H_\text{gw}^\text{L},H_\text{em}^\text{det})$, and $Z_k^\mathrm{det} = p(d_\mathrm{gw}| d_\mathrm{em},H^k,H_\mathrm{gw}^\mathrm{L})$.

Therefore, we can approximate full odds $O^i$ with $O_\text{det}^i$.

We apply Laplace approximation to simplify the estimation of $O^i_\text{det}$. For the posterior which can be well-approximated with Gaussian likelihood in $k$ parameters,
\begin{equation}
 Z_i\approx L_i(\widehat{\boldsymbol\eta}_i)
 \pi_i(\widehat{\boldsymbol\eta}_i)(2\pi)^{k/2}|\Sigma_i|^{1/2},
 \label{eq:baseline_laplace}
\end{equation}
where $\widehat{\boldsymbol\eta}_i$ is the mode and $\Sigma_i$ is the local
covariance. For a uniform prior of volume $V_i$, the factor multiplying the
maximum likelihood is $(2\pi)^{k/2}|\Sigma_i|^{1/2}/V_i$.
We set prior volumes for each sampling runs on different cluster lens to be the same. Therefore,
\begin{equation}
 \widetilde O_{\rm det}^i
 =\frac{\pi_i\exp(\ell_i^*)}{\sum_{j\ne i}\pi_j\exp(\ell_j^*)},
 \qquad
 \ell_i^*=\max_{\boldsymbol\eta}\log L_i(d\mid\boldsymbol\eta),
 \label{eq:baseline_peak_odds}
\end{equation}
which reduces to the main-text expression for equal cluster priors.

\section{Calculation of Deflection Angle and Uncertainty from the Numerical Models}
\label{app:maps}

The values of $D_s$ and $D_{ls}$ depend on the source redshift and the Hubble constant \citep{IntroToCosmology2015}:

\begin{equation}
D_A = \frac{1}{1+z_s}\int_{0}^{z_s}\frac{c}{H(z')}dz'\:, \:\: H(z)=H_0\sqrt{\Omega_m(1+z)^3+\Omega_\Lambda},
\end{equation}

where $D_A$ is the angular diameter distance, and $\Omega_m$ and $\Omega_\Lambda$ are the matter and dark energy density parameters respectively. Flat $\Lambda$CDM cosmology is assumed in this study, with $\Omega_m=0.3$ and $\Omega_\Lambda=0.7$ for simplicity.

Also, the data in 2D deflection maps are reduced deflection vector:

\[\vec{\alpha}_{\text{map}} (\vec{\theta}) \equiv (\frac{D_{ls}}{D_s})_{\text{map}}\,\hat{\alpha}(D_l\vec{\theta})=\hat{\alpha}(D_l\vec{\theta}).\]

Thus, the actual deflection angle can be obtained as

\begin{equation}\vec{\alpha}_{\text{real}} (\vec{\theta}) =(\frac{D_{ls}}{D_s})_{\text{real}}\,\vec{\alpha}_{\text{map}} (\vec{\theta}).\end{equation}

The corresponding image positions can then be obtained using $\vec{\alpha}_{\text{real}} (\vec{\theta})$.

The angular diameter distances depend on $z_s$. For $H_0$, however, the dependence cancels out in the ratio $D_{ls}/D_s$. Therefore, the deflection maps can be scaled directly without requiring the precise value of $H_0$.

The uncertainty of $\vec{\alpha}_{\text{obs}} (\vec{\theta})$ is evaluated using a set of simulated lens potential maps from simulations of numerical lens models. To obtain a rough estimate of the uncertainty in the deflection angle, we first obtain the set of deflection angle maps by simple differentiation ($\vec{\alpha}(\vec{\theta}) = \nabla \psi (\vec{\theta})$). We then compute the standard deviation of the deflection angle at each pixel ($|\alpha|^2=\alpha_x^2+\alpha_y^2$) across the simulated maps and adopt the upper $+1\sigma$ percentile as the scalar uncertainty value $\sigma_{|\alpha|}$. The uncertainty of lens potential $\sigma_{\psi}$ is also obtained by the similar approach with the lens potential map directly.

We then calculate the uncertainty of time delays by error propagation:

\begin{equation}
    \sigma_{\tau} = D_{\Delta \tau}\sqrt{|\alpha|^2 \sigma_{|\alpha|}^2 + \sigma_{\psi}^2}.
\end{equation}

\begin{figure*}
    \includegraphics[width = \textwidth]{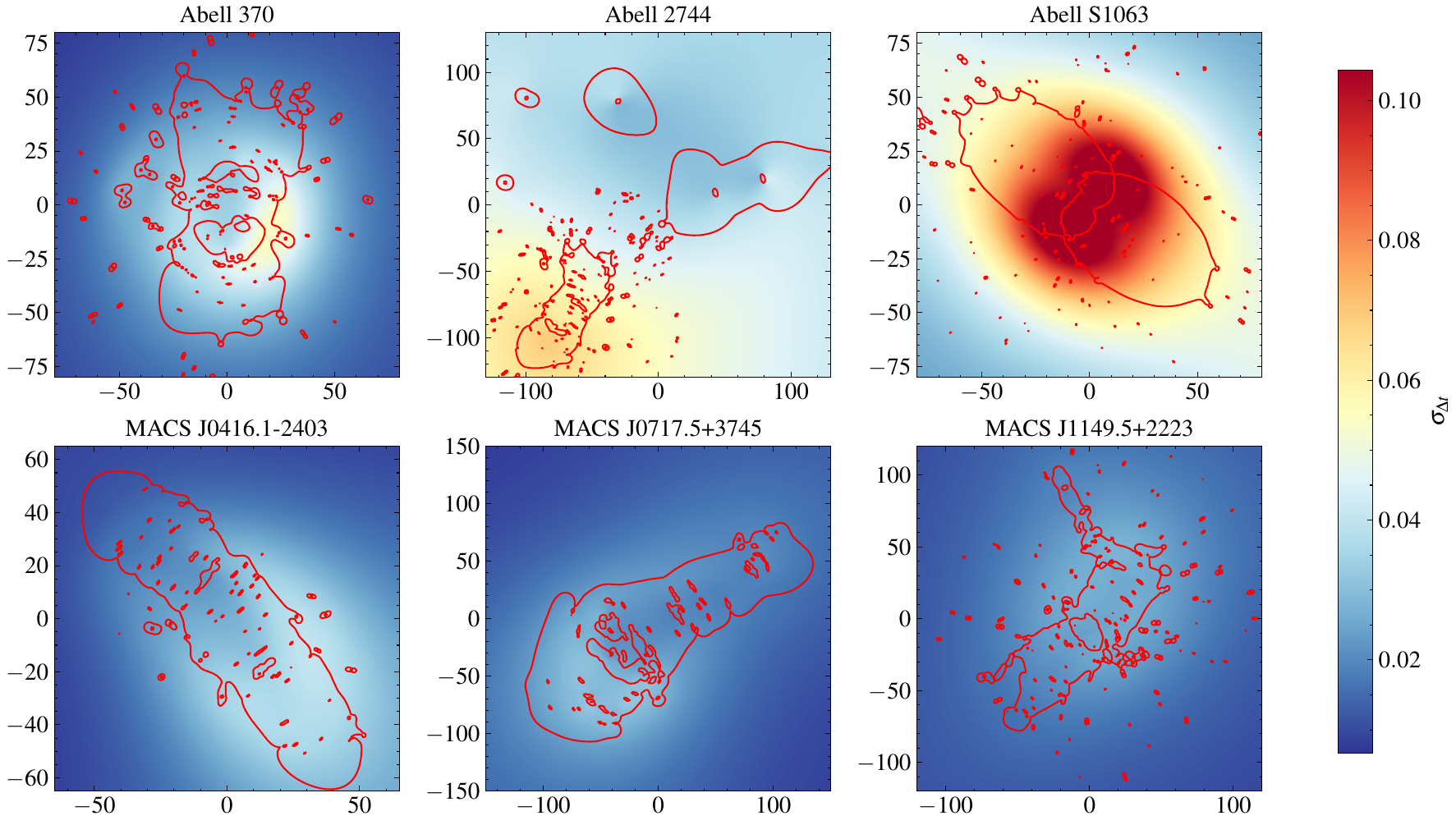}
    \caption{
    $\sigma_{\Delta \tau}$ maps of the 6 galaxy clusters used in this study. The color bar represents the relative uncertainty on time delay measurements for each pixel, and red curves are critical curves for each cluster lens. In the core of the galaxy clusters, the relative uncertainties are higher and can reach up to 12\%. Therefore, sources in this study which are behind core region of cluster lens are subjected to higher uncertainties of time delays of images.}
    \label{fig:sigma_tau}
\end{figure*}

As relative time-delay is used in the study, the uncertainty is 

\begin{equation}
    \sigma_{\Delta\tau}^2 = \sigma_{\tau_i}^2 + \sigma_{\tau_0}^2,
\end{equation}

where $\sigma_{\tau_i}$ is the uncertainty of time delay of image i and $\sigma_{\tau_0}$ is the uncertainty of time delay of the first image detected. For simplicity, the upper bound of the possible values are used. The resulting $\sigma_{\Delta \tau}$ maps of the lens models used are shown in Fig.~\ref{fig:sigma_tau}. We then obtain $\sigma_{\Delta \tau}$ for each
pixel of the numerical lens model and obtain the value of a
given image position by linear interpolation.

\section{Approximation of Likelihood for Localization}
\label{app:local}

For this study, we assume Gaussian noise in the detection of signals, which is a standard approach. By modeling the observed data as $d_i = n_i + s_i$, where $d_i$ is the observation, $n_i$ is the noise, and $s_i$ is the true signal, the likelihood of the data given the signal is:

\begin{equation}
    p(d_i| s_i) = p(n_i = d_i - s_i) \sim \text{Gaussian}(\mu = s_i, \sigma^2).
\end{equation}

Therefore, set $s_i = f(\beta, z_s, d_\text{em}^i, H^i, H_\text{gw}^\text{L})$ and $d_i$ as GW signals, the likelihood becomes:

\[p(d_\text{gw}|\beta,z_s,d_\text{em}^i,H^i,H_\text{gw}^\text{L})\sim \text{Gaussian},\]
and
\begin{equation}
\begin{split}
   p(d_\text{gw}|\beta,z_s,d_\text{em}^i,H^i,H_\text{gw}^\text{L})
  &= \prod_{j=1}^{N}\frac{1}{\sqrt{2\pi}\,\sigma_{\DLeff}}
    \exp\!\biggl[-\frac{(\DLeff^{\text{obs},j}-\DLeff^{j})^2}{2\sigma_{\DLeff}^2}\biggr] \cdot \\
  &\quad \prod_{k=1}^{N-1}
    \frac{1}{\sqrt{2\pi}\,\sigma_{\Delta\tau}}
    \exp\!\biggl[-\frac{(\Delta \tau_{k,\text{obs}}-\Delta \tau_{k,\text{H}})^2}
    {2\sigma_{\Delta\tau}^2}\biggr].
\end{split}
\end{equation}

Note that we assume $\sigma_{\Delta \tau}\approx\sigma_{\Delta \tau,\text{lensmodel}}$ and

$$
\sigma^2_{D_L}=\sigma^2_{D_L,\text{GW}}+\sigma^2_{D_L,\text{lensmodel}},
$$

since $\sigma_{\Delta \tau,\text{GW}}$ is much smaller than $\sigma_{\Delta \tau,\text{lensmodel}}$.

\bibliography{sample701}{}
\bibliographystyle{aasjournalv7}
\end{document}